\documentclass[a4paper,11pt,dvips]{article}
\usepackage{graphicx}
\usepackage{amsmath}
\usepackage{amssymb}
\usepackage{latexsym}

\newcommand{\bra}{\begin{array}}
\newcommand{\era}{\end{array}}
\newcommand{\beq}{\begin{equation}}
\newcommand{\eeq}{\end{equation}}
\newcommand{\beqar}{\begin{eqnarray}}
\newcommand{\eeqar}{\end{eqnarray}}

\def\BC{\bb C}
\def\_\BC{\bbi C}

\def\bz {\bar{z}}

\def\( {\left(}
   \def\) {\right)}
\def\[ {\left[}
\def\] {\right]}
\def\no2 {{\textstyle{n\over 2}}}

\newcommand{\om}{\omega}

\newcommand{\si}{\sigma}

\newcommand{\te}{\theta}

\newcommand{\pa}{\partial}

\newcommand{\lga}{\longrightarrow}

\newcommand{\da}{\dagger}

\newcommand{\sq}{\sqrt}

\newcommand{\lb}{\label}

\newcommand{\PRL}[1]{ {\it Phys.~Rev.~Lett.} {\bf #1}}

\newcommand{\JP}[1]{ {\it J.~Phys.} {\bf #1}:\  Math.~Gen.~}

\newcommand{\JMP}[1]{ {\it J. Math.~Phys.} {\bf #1}}

\begin{document}
\thispagestyle{empty}

\begin{flushright}
ucd-tpg/06-05\\
hep-th/0611301
\end{flushright}

\vspace{0.5cm}
\begin{center}
 {\Large \bf   Noncommutative Description  of  Spin Hall Effect }\\

\vspace{0.5cm}

{\bf Ahmed Jellal$^{a,b,}$\footnote{E-mail : ajellal@ictp.it -- jellal.a@ucd.ac.ma}}
and {\bf Rachid Hou\c ca$^{b,}$\footnote{E-mail : rachid.houca@gmail.com}}\\
\vspace{0.5cm}

{\em $^a$Center for Advanced Mathematical Sciences,
College Hall,
American University of Beirut,\\
P.O. Box 11-0236, Beirut, Lebanon }

{\it  $^b$Theoretical Physics Group, Laboratory of Condensed Matter Physics,
Faculty of Sciences, \\ Choua\"ib Doukkali University,
P.O. Box 4056,
24000 El Jadida, Morocco}\\[1em]

\vspace{3cm}
{\bf Abstract}
\end{center}
\baselineskip=18pt
\medskip

We propose an approach based on
a generalized quantum mechanics
to deal with the basic features of the
intrinsic spin Hall effect. This can be done by
considering two decoupled harmonic
oscillators on the noncommutative plane and evaluating the
spin Hall conductivity. Focusing on the high frequency regime,
we obtain a diagonalized Hamiltonian. After getting the corresponding
spectrum, we show that there is a Hall conductivity without
an external magnetic field,
which is noncommutativity
parameter $\te$-dependent. This allows us to make contact with
the spin Hall effect and also give different interpretations.
Fixing
$\te$, one can recover three different approaches
dealing with  the phenomenon.


\newpage

\section{Introduction}

The spin Hall effect (SHE) is physical phenomenon, which
has been discovered in $1971$ by D'yakonov and
Perel~\cite{perel}. It is a consequence of
the
spin-orbit coupling where an applied
electric field to a sample can lead to
a spin transport in perpendicular direction and
spin accumulation at the lateral edges~\cite{kato,wunderlich}.
It is characterized by a spin Hall conductivity resulting from the
spin polarization on the boundaries of the sample.
There are two types of SHE: intrinsic and extrinsic,
each one is depending to what kind of spin-orbit coupling
contribution to the considered Hamiltonian
describing the system~\cite{rashba}.

The intrinsic SHE has been
theoretically predicted for semiconductors
with spin-orbit interactions. 
Indeed,  Sinova {\it et al.}~\cite{sinova}
described a new effect in  $n$-type semiconductor spintronics that leads to
dissipationless spin-currents in paramagnetic spin-orbit coupled systems.
They argued that in a high mobility two-dimensional electron system with
substantial Rashba spin-orbit coupling, a spin-current that flows
perpendicular to the charge current is intrinsic. In the usual case
where both spin-orbit split bands are occupied, the spin-Hall
conductivity has a universal value.
Other related works can be found in references~\cite{sinova2,shen,hu}.

The theoretical prediction of the intrinsic SHE has been also
argued by another group~\cite{murakami1}. This has been done by
adopting a mathematical formalism governed
by the Luthinger Hamiltonian for $p$-type semiconductors in
two-dimensions. In fact,
Murakami
{\it et al.}~\cite{murakami1}
showed that the electric field can generate a
dissipationless quantum spin current at room temperature, in hole
doped semiconductors such as {\it Si, Ge} and {\it GaAs}.
Taking advantage of a generalization
of the quantum Hall effect~\cite{prange} to higher
dimensional manifolds, they showed that  the intrinsic SHE leads to efficient
spin injection without the need for metallic ferromagnets. Another derivation
has been established by using the Berry phase approach,
which can be found in~\cite{murakami2}.

Very recently, a quantum version of the intrinsic
SHE has been reported
by Bernevig and Zhang~\cite{zhang}. 
In fact, by
considering a Hamiltonian brought form solid state physics,
they showed that the spin Hall conductivity is quantized in
units of $e\over 2\pi$ and built the corresponding wavefunctions.
These have strong overlapping with those have been construct by Halperin
many years ago~\cite{halperin} or
their equivalents in terms of the matrix model
theory~\cite{jel}. These latter have been formulated to
describe
the quantum Hall effect generated from charged
particles by treating theirs spins as additional degrees
of freedom.

Based on the above works and in particular~\cite{zhang}, we
describe our main idea.
More precisely, we quantum mechanically
develop another approach to analyze the intrinsic
SHE. This can be done
by resorting the spectrum of two noncommutative harmonic oscillators and
evaluating the spin Hall conductivity. Solving the Hamiltonian system at
high frequency regime, we derive the corresponding eigenvalues as well as
eigenstates. Using these to get the Hall conductivity of charge without
an external
magnetic field and therefore the spin Hall conductivity, which are noncommutativity
parameter $\te$-dependents. Since $\te$ is
a free parameter, one can differently
interpret our results. Indeed, for some particular values of $\te$
we discuss how to get the
 quantum SHE
by constructing the Laughlin
wavefunction analogue. Furthermore,
we establish a link between our approach and those
proposed by
 Bernevig and Zhang, Sinova {\it et al.} and Murakami {\it et al.}

The present paper is organized as follows. In section 2,
we consider two decoupled harmonic oscillators on the noncommutative
plane $\mathbb{R}^2_{\te}$
and getting its spectrum at high frequency regime. This allows us
to make contact with
the Landau problem on the ordinary plane $\mathbb{R}^2$ and therefore
Laughlin wavefunctions at the
filling factor $\nu={1\over m}$~\cite{laughlin}, with $m$ is an odd integer.
In section 3, to determine the spin Hall conductivity, we introduce the electric
field through a confining potential resulting from our
consideration. For this, we distinguish two cases:
spin up and down, which are relatively found to be equivalents up to a minus sign.
In section 4, we offer different interpretations of our results
by showing how some theories on the subject can be recovered from
our analysis. We conclude and give
some perspectives in the last section.

\section{ Two oncommutative harmonic oscillators}
We consider two decoupled harmonic oscillators on $\mathbb{R}^2_{\te}$
and determine the corresponding eigenvalues as well as
eigenfunctions. This can
be done by introducing the star product and the ordinary commutation relations in
quantum mechanics. Restricting to the high frequency regime, we obtain a
diagonalized Hamiltonian as well as its spectrum.
We give different comparisons with respect to the
Landau problem on $\mathbb{R}^2$  in order to show its overlapping
with our approach.

\subsection{Hamiltonian of system}

Our proposal can be elaborated by considering
two decoupled harmonic
oscillators of the same masses $m$
and frequencies $\om$ on $\mathbb{R}^2$. They are described by the Hamiltonian
\beq \lb{ccc}
H_{\sf plane}={1\over 2m}
\left(p_{x}^{2 }+p_{y}^{2}\right) +{m\om^2\over 2}\left(x^2+y^2\right)
\eeq
which can be interpreted as a Hamiltonian for one-particle system on
$\mathbb{R}^2$  in absence of any interacting term. 
 It can be diagonalized by introducing
the creation and annihilation operators
\beq
a_i= {1\over\sqrt{2\hbar m\om}}p_i -i \sqrt{m\om\over 2\hbar},
\qquad a_i^{\da}= {1\over\sqrt{2\hbar m\om}}p_i +i \sqrt{m\om\over 2\hbar},
\qquad i=x,y
\eeq
where the only non-vanishing commutator is
\beq
\left[a_i, a_i^{\da} \right]= \mathbb{I}.
\eeq
These are implying that $H_{\sf plane}$ can be arranged as
\beq
H_{\sf plane}={\hbar \om \over 2}\left(a^{\da}_xa_x+a^{\da}_ya_y+1\right)
\eeq
where the corresponding eigenstates are
\beq
|n_x,n_y\rangle={(a^{\dagger}_x)^{n_x}\over
  \sq{n_x!}}{(a^{\dagger}_y)^{n_y}\over \sq{n_y!}}\ |0,0\rangle
\eeq
as well as the eigenvalues
\beq
E_{n_x,n_y}={\hbar \om \over 2}\left(n_x+n_y+1\right), \qquad n_i = 0,1,2,\cdots
\eeq
where
$|0,0\rangle$ is the fundamental state.
Next, we will see how these results can be generalized to
$\mathbb{R}^2_{\te}$  and used to deal with our issues.

In doing our generalization, we adopt a method similar to
that used in~\cite{jellal}. Indeed,
the canonical quantization of the system described by (\ref{ccc})
is achieved by
introducing the coordinate $r_j$ and momentum $p_k$ operators
satisfying the relation
\beq \lb{ddd}
\left[r_j,p_k\right]=i\hbar \delta_{jk}.
\eeq
But to deal with our proposal,  we consider a generalized
quantum mechanics governed by
(\ref{ddd}) and the noncommutative coordinates, such as
\beq \lb{aaa}
\left[x,y\right]=i\te
\eeq
where $\te$ is a real free parameter and has length square
of dimension.
Without loss of generality, hereafter we assume that $\te>0$.
Noncommutativity can be imposed
by treating the coordinates as commuting but requiring that
composition of their functions is given in terms of the star
product
\beq \lb{abc}
\star \equiv \exp{i\te\over 2}\left(\overleftarrow {\pa_x}\overrightarrow{\pa_y}-
\overleftarrow{\pa_y}\overrightarrow{\pa_x}\right).
\eeq
Now, we deal with the commutative coordinates $x$ and $y$ but replace the
ordinary products with the star product~(\ref{abc}). For example, instead of the
commutator~(\ref{aaa}) one defines
\beq
x\star y-y\star x=i\te.
\eeq
At this level,
let us  derive the corresponding form of the Hamiltonian
(\ref{ccc}) in terms of the noncommutative
coordinates~(\ref{aaa}).
First, we quantize the present system by establishing the
commutation relation~(\ref{ddd}). Second, we take into account
the noncommutativity of the coordinates
by defining a new operator as
\beq
H\star \psi(\vec {r})\equiv H^{\sf nc}\psi(\vec {r})
\eeq
where $\psi(\vec {r})$ is an arbitrary eigenfunction of $H$.
By doing this processing, we obtain
the noncommutative version of the Hamiltonian (\ref{ccc}).
This is
\beq \lb{r}
H^{\sf nc}=\left[{1\over 2m}+{m\om^2\over 2}\left({\te\over
  2\hbar}\right)^2\right]\left(p_{x}^{2}+p_{y}^{2}\right)
+{m\om^2\over 2}\left(x^2+y^2\right)
+{m\om^2\te\over 2\hbar} \left(yp_x-xp_y\right).
\eeq
We emphasis that
due to the noncommutativity between spacial coordinates we ended up with
two coupled harmonic oscillators. This coupling is described
in terms of the angular momenta $L_z(\te)$
\beq\lb{amo}
L_z(\te) ={m\om^2\te\over 2\hbar} \left(yp_x-xp_y\right).
\eeq
It is obvious that (\ref{amo}) disappears once we set $\te=0$ and then
recover (\ref{ccc}).
 $L_z(\te)$ is analogue to that
corresponding to the Landau problem on ${\mathbb R}^2$,
see next.

\subsection{High frequency regime}
The Hamiltonian $H^{\sf nc}$ can not be diagonalized
directly, we need to introduce
some relevant approximation in order to get its spectrum.
For this, we restrict ourselves to the high
frequency ($\sf hf$) regime, which is characterized by
the limit
\beq\lb{hf}
\left[ {1\over 2m}+ {m\om^2\over
  2}\left(\te\over2\hbar\right)^2\right]\simeq
{m\om^2\over 2}{\left(\te\over 2\hbar\right)}^2.
\eeq
This is not surprising, since an analogue approximation
has been employed by Berniveg and Zhang~\cite{zhang}
in analyzing the quantum version of
the intrinsic SHE on ${\mathbb R}^2$. We will be back to
clarify this point in section 4.
In the limit~(\ref{hf}), the Hamiltonian (\ref{r}) reduces to
\beq\lb{hfh}
H_{\sf hf}(\te)={m\om^2\over 2}\left[{\left(\te\over 2\hbar\right)^2}
(p_{x}^2+p_{y}^2)+x^2+y^2+{\te\over 2\hbar}(yp_{x}-xp_{y})\right].
\eeq
Let us give a comment about our Hamiltonian.
It is interesting to note that (\ref{hfh}) has a
strong overlapping with the Landau problem on ${\mathbb R}^2$. To see this,
we start by recalling that in the symmetric gauge
\beq
A={B\over 2}\left(y,-x\right)
\eeq
the Landau Hamiltonian for a one-charged particle of
mass $m$ in two-dimensions and submitted to
an uniform magnetic field $B$ is
given by
\beq\lb{laham}
H_{\sf landau} = {1\over 2m}\left[
\left(p_{x}^2+p_{y}^2\right)+ \left({eB\over 2c}\right)^2\left(x^2+y^2\right)
+{eB\over c}(yp_{x}-xp_{y})\right].
\eeq
Clearly $H_{\sf landau}$ is sharing some common features with $H_{\sf hf}$. This can
be shown by
requiring that the conditions
\beq\lb{teb}
\te_{\sf landau} ={2\hbar c\over eB}=2l_B^2, \qquad 2\om =\om_c
\eeq
is fulfilled
where $l_B$ is the magnetic length. Therefore, one may
interpret $\te$ as an external parameter $B$ applied to
the system,
which remains among the important values of $\te$ derived right now.
Consequently,
since  (\ref{laham}) is the cornerstone of
the quantum Hall effect~\cite{prange},
then
$H_{\sf hf}(\te)$ will allows us to make
contact with this  effect.

In the forthcoming analysis,
it is convenient to consider the complex plane $(z,\bz)$ where $z=x+iy$
and $p_z={1\over 2}\left(p_x -i p_y\right)$.
In this case,
(\ref{c}) can be written as
\beq\lb{hamz}
H_{\sf hf}(\te)={m\om^2\over 2}\left[4{\left(\te\over 2\hbar\right)^2}
p_{z}p_{\bz} + z\bz+{\te\over 2\hbar}(zp_{z}-\bz p_{\bz})\right].
\eeq
As usual the diagonalization of (\ref{hamz}) can be realized by
introducing the creation and annihilation operators. They are given by
\beq
a={\sqrt{\te}\over \hbar}p_{\bz} -{i\over 2\sqrt{\te}}z,\qquad
a^{\dagger}={\sqrt{\te}\over \hbar}p_{z} +{i\over 2\sqrt{\te}}\bz.
\eeq
It easy to show that
\beq
[a,a^{\dagger}]={\mathbb I}
\eeq
and other commutators are nulls. In terms of $a$ and $a^{\dagger}$,
$H_{\sf hf}(\te)$ can be mapped as
\beq\lb{aah}
H_{\sf hf}(\te)={m\om^2\te\over 4\hbar}\left(2a^{\dagger}a+1\right).
\eeq
This is nothing but one-dimensional harmonic oscillator with frequency
\beq
\om(\te)={m \om^2\te \over  \hbar^2}.
\eeq
The corresponding
 spectrum  can be obtained
by solving the eigenvalue equation
\beq
H_{\sf hf}(\te)\phi=E_{ n}(\te)\phi
\eeq
to get the eigenfunctions
\beq\lb{1ewfp}
\phi_{n}(z,\bz,\te)={z^n}
\exp\left(-{z\bz\over 2\te}\right)
\eeq
upon a factor of normalization. The associated   energy levels
are given by
\beq\lb{1elp}
 E_{n}(\te)={m\om^2\te\over 4\hbar}\left(2n+1\right), \quad
 n=0,1,2, \cdots.
\eeq

The previous analysis can be generalized to a system of
$N$-identical particles governed by the total Hamiltonian
\begin{equation}\lb{nhp}
H_{\sf hf}^{\sf tot} (\te)={m\om^2\over 2}\sum_{i=1}^N
\left[4{\left(\te\over 2\hbar\right)^2}
p_{z_i}p_{\bz_i} + z_i\bz_i+{\te\over 2\hbar}(z_ip_{z_i}-
\bz_i p_{\bz_i})\right].
\end{equation}
where the total energy is $N$-copies of (\ref{1elp}) and the eigenvalues
is basically the tensorial product of $N$ those given in (\ref {1ewfp}).
If the system is living on  the lowest level,
which of course means that all $n_i=0$
with $i=1,\cdots,N$ and each $n_i$ corresponds to the
spectrum~(\ref{1ewfp}--\ref{1elp}),
the total wavefunction
can be written in terms of the Vandermonde determinant. This is
\beq\lb{1ls}
\Phi^1_{\sf hol}(z,\bz,\te)=\prod_{i<j}(z_i-z_j)
\exp\left(-{1\over 2 \te}\sum_i |z_i|^2\right)
\eeq
which is an holomorphic wavefunction.
It is obvious that by using the constraint
(\ref{teb}), one can recover the Laughlin wavefunction at
the filling factor $\nu=1$~\cite{laughlin} describing
charged particles
in the presence of an uniform magnetic field.
Therefore, (\ref{1ls}) can be interpreted as
the Laughlin wavefunction
 analogue
and other similar ones  can be constructed
as
\beq\lb{mls}
\Phi^m_{\sf hol}(z,\bz,\te)=
\prod_{i<j}(z_i-z_j)^m\exp\left(-{1\over 2\te}\sum_i |z_i|^2\right)
\eeq
with $\nu={1\over m}$ and $m$ has odd integer values.

Later, we will see how the above  results
can be employed to deal with the
basic features of the intrinsic SHE. In fact, we show that (\ref{aah}) could lead to
 the spin Hall  conductivity comparable with those derived by other groups.

\section{ Spin Hall conductivity}
Before evaluating the spin Hall conductivity, let us emphasis
an important point. Through the present analysis, we
are considering  electrons of spin $1\over
2$. Thus, we need to distinguish two possible configurations:
spin up and down cases. Consequently, to reflect the
spin--orbit coupling contribution, we should have two Hamiltonians differing
between each other by a sign of the angular momenta
term~(\ref{amo}). To reproduce this effect,
we simply identify spin up to the noncommutativity parameter
$+\te$ and spin down to $-\te$. Subsequently, we analyze
each case by establishing all ingredients to
show that our system is really exhibiting an intrinsic SHE.

\subsection{ Electric field components}

There is an important ingredient that should be fixed before talking
about the intrinsic SHE. This is the electric field, which is
responsible of having such phenomenon.
More precisely, an electric current
passes through a system with spin-orbit coupling, induces a spin
polarization near the lateral edges. This leads to a spin
accumulation and therefore a spin Hall conductivity.
For this, we will show how to fix the external parameter
in our approach.

To reproduce the required field in terms of our language,
we can simply use the  standard definition, which is showing
that
\beq
\vec{F}=-e\vec{E}=-\overrightarrow{\mbox{ grad}} \ V
\eeq
where the scalar potential
$V$ can be derived from
(\ref{c}). It follows that
$V$ should be nothing but a
confining potential, such as
\beq
V={m\om^2\over 2}\left(x^2+y^2\right).
\eeq
Combing all to get the electric field components
\beq \lb{111}
E_{x}={m\om^2\over e}x, \qquad
E_y={m\om^2\over e}y.
\eeq
As we will see later, analogue relations to (\ref{111}) have be
introduced by Berniveg and Zhang in analyzing
the quantum SHE.
Consequently, (\ref{111})
will play in crucial role in dealing with
the subject. To clarify this point, let us treat separately
spin up and down cases.

\subsection{ Spin up case}

As we claimed before, the spin up case can be identified
to $+\te$. Therefore, this case
is describing by the Hamiltonian
(\ref{c}) as well as its corresponding
analysis reported before.
To determine the spin Hall conductivity
for spin up,
we start by evaluating the velocity components
to get first the Hall conductivity of charge
and second return to deal with our issues.

Let us begin by determining the velocity component
 along $x$-direction. It
can be obtained by using the Heisenberg
equation
\beq
v_x(+\te)={i\over \hbar}\left[H_{\sf hf}(+\te),x\right].
\eeq
To
derive the Hall current of charge, we need to calculate the expectation
value of $v_x(+\te)$. This can be done with respect to the eigenstates
$\phi_n(z,\bz,+\te)$ to get
\beq
\langle v_x(+\te)\rangle={m\om^2\te \over 2\hbar}y.
\eeq
The relation between velocity and current implies
\beq \lb{222}
\langle j_x(+\te)\rangle={\rho em\om^2\te\over 2\hbar}y
\eeq
where $\rho ={N\over S}$ is the particle density
and $S$ is the system surface. Now, we have all ingredients to
derive the Hall conductivity of charge. Indeed, using
the second relation in (\ref{111}), we obtain
\beq
\sigma_{xy}(+\te)={\rho e^2\over 2\hbar}\te.
\eeq
It is interesting to note that unlike the Landau problem,
we have a transversal conductivity
without an external  magnetic
field $B$.
This shows that our system can be seen as a Hall system
and then can be used to establish another approach dealing with
the basic features of the quantum Hall
effect.
To recover, the Landau problem study we simply
 identify $\te$ to $B$ through the relation (\ref{teb}).

Using the same analysis as before, we show that the
Hall current along $y$-direction is given by
\beq
\langle j_y(+\te) \rangle=-{\rho em\om^2\te\over 2\hbar}x
\eeq
and therefore the Hall conductivity $\sigma_{yx}(\te)$ is
\beq
\sigma_{yx}(+\te)=-{\rho e^2\over 2\hbar}\te.
\eeq
It is clear that
\beq\lb{pmis}
\sigma_{xy}(+\te)=-\sigma_{yx}(+\te)
\eeq
as it is well-known in the quantum Hall effect world. For this,
we only focus on $\sigma_{xy}(+\te)$ in the forthcoming
analysis.

Up to now we have derived the Hall conductivity of charge, which
basically came from a
deformation of the space ${\mathbb R}^2$. It is natural to ask about the spin Hall
conductivity $\sigma_{xy}^s (+\te)$. Indeed,
since an electron with charge $e$  carries a spin ${\hbar\over 2}$, a
factor of $\hbar\over 2e$ is used to convert the charge
conductivity into the spin conductivity~\cite{zhang}.
Applying this statement to our case, we should have
\beq
\sigma_{xy}^s (+\te)=2\sigma_{xy}(+\te){\hbar\over 2e}.
\eeq
Finally, the spin Hall conductivity is
\beq \lb{SHC}
\sigma_{xy}^s(+\te)={\rho e\over 2}\te
\eeq
which is noncommutativity parameter
$\te$-dependent and represents
the main result derived so far in the present paper.
Note that, once $\te$ is swished off $\sigma_{xy}^s(+\te)$ goes to zero.
Later, we will see how it can be used to offer
different interpretations.

\subsection {Spin down case}

To accomplish our analysis, we consider the second part of electron
that is the spin down case. This is corresponding to
change $+\te$ by $-\te$ in the above study. Otherwise, it is equivalent to
consider the commutator
\beq\lb{-te}
\left[x,y\right]=-i\te
\eeq
instead of its analogue given by (\ref{aaa}).
Using the same analysis as before, we end up with a
Hamiltonian describing two harmonic oscillators on $\mathbb{R}^2_{\te}$
generated by (\ref{-te}). In fact,
at high frequency regime, it is given by
\beq\lb{sdown}
H_{\sf hf}(-\te)={m\om^2\over 2}\left[{\left(\te\over 2\hbar\right)^2}
(p_{x}^2+p_{y}^2)+x^2+y^2-{\te\over 2\hbar}(yp_{x}-xp_{y})\right]
\eeq
which is analogue to that describing
one-electron of spin down that has been considered
by Berniveg and Zhang \cite{zhang}, see later.

At this stage, let us exhibit the effective spin--orbit coupling
 in our formalism. In doing so, we can use
 (\ref{hf}) and  (\ref{sdown})
together to define a total Hamiltonian as
\beq\lb{totham}
H_{\sf hf}^{\sf tot} =
\left(
  \begin{array}{cc}
   H_{\sf hf}(+\te) & 0 \\
   0 & H_{\sf hf}(-\te) \\
  \end{array}
\right).
\eeq
Equivalently, one can write
\beq\lb{pmham}
H_{\sf hf}^{\sf tot}= {m\om^2\over 2}\left[{\left(\te\over 2\hbar\right)^2}
(p_{x}^2+p_{y}^2)+x^2+y^2 +{\te\over 2\hbar}
\left(
  \begin{array}{cc}
   1 & 0 \\
   0 & -1 \\
  \end{array}
\right)
(yp_{x}-xp_{y})\right]
\eeq
where the third component of spin is
\beq
S_z ={\hbar\over 2} \left(
  \begin{array}{cc}
   1 & 0 \\
   0 & -1 \\
  \end{array}
\right).
\eeq
Therefore,
the last term in (\ref{pmham}) is resulting from an effective
interaction where $\te$ is playing the role of a coupling parameter.
It is obvious that
The derived interaction disappears if $\te$ is switched off.

In similar way to spin up, $H_{\sf hf}(-\te)$ can be diagonalized by
setting the creation and annihilation operators. They are
\beq\lb{bop}
b={\sqrt{\te}\over \hbar}p_{\bz} +{i\over 2\sqrt{\te}}z,\qquad
b^{\dagger}={\sqrt{\te}\over \hbar}p_z -{i\over 2\sqrt{\te}}\bz\qquad
\eeq
which satisfy the relation
\beq
\left[b ,b^{\dagger} \right] = {\mathbb I}.
\eeq
With these, (\ref{sdown})  reads as
\beq
H_{\sf hf}(-\te)={m\om^2\te\over 4\hbar}\left(2b^+b+1\right).
\eeq
This Hamiltonian has the same form as that for spin up, but
the main difference is that the corresponding eigenfunctions are antiholomorphic,
such as
\beq\lb{anf}
\phi_{k}(z,\bz,-\te)={(\bz)^k}
\exp\left(-{z\bz\over 2\te}\right)
\eeq
up on a factor of normalization
and the energy levels are given by
\beq\lb{ans}
E_k(-\te)={m\om^2\te\over 4\hbar}(2k+1), \qquad k=0,1,2\cdots.
\eeq

The above results for one-electron of spin down can be generalized to
a system of $N$-identical electrons of spin down. In particular, for
$N$-particles in the lowest level, namely  $k_i=0$
with $i~=~1,\cdots,N$ and each $k_i$ corresponds to the spectrum~(\ref{anf}--\ref{ans}),
the total wavefunction is
\beq\lb{an1ls}
\Phi^1_{\sf anti}(z,\bz,-\te)=\prod_{i<j}(\bz_i-\bz_j)\exp\left(-{1\over 2\te}\sum_i |z_i|^2\right)
\eeq
which is antiholomorphic and analogue to the first Laughlin wavefunction at
$\nu=1$. Other analogue wavefunctions can be written as
\beq\lb{anmls}
\Phi^m_{\sf anti}(z,\bz,-\te)=
\prod_{i<j}(\bz_i-\bz_j)^m\exp\left(-{1\over 2\te}\sum_i |z_i|^2\right).
\eeq
These as well as their holomorphic parters
 (\ref{mls}) will be used to built
the whole wavefunctions describing two sectors
where each one contains $N$-electrons of
spin up or down.

As before the spin Hall
conductivity  $\sigma_{xy}^s(-\te)$ corresponding to $H_{\sf hf}(-\te)$
can be calculated by using the Heisenberg equation
\beq
v_x(-\te)={i\over \hbar}\left[H_{\sf hf}(-\te),x\right].
\eeq
This shows that the  Hall conductivity for charge is
\beq
\sigma_{xy}(-\te)=-{\rho e^2\te\over 2\hbar}.
\eeq
Using the same statement as for spin up to obtain
the $\sigma_{xy}^s(-\te)$ resulting from $N$-electrons of spin down
in the presence of an electric field $E_x$ (\ref{111}). This is
\beq \lb{shcmt}
\sigma_{xy}^s(-\te)=-{\rho e\over 2}\te.
\eeq
Similarly along  $y$-direction, we have
\beq\lb{anSHC}
\sigma^s_{yx}(-\te)={\rho e\over 2}\te.
\eeq
Combining all, we can arrange all
conductivities for the $x$-direction as
\beq
\sigma^s_{xy}=\left\{
\begin{array}{lll}
\ \ {\rho e\over 2}\te \qquad {\mbox{for\ (\ref{aaa})}}\\
\ \ 0\qquad \  \ \ {\mbox{for\  $\te=0$}}\\
-{\rho e\over 2}\te \qquad {\mbox{for \ (\ref{-te})}}
\end{array}\right.
\eeq
where  (\ref{aaa}) and (\ref{-te}) are two different deformations
of plane introduced to reflect an effective spin--orbit coupling
contribution to the Hamiltonian (\ref{ccc}).
Similar equation, up to a minus sign, can be derived for
$\sigma^s_{yx}$ along $y$-direction.
By comparing (\ref{SHC}) and (\ref{anSHC}), it follows that
the constraint
\beq\lb{pmte}
\sigma_{xy}^s(+\te) = -\sigma_{xy}^s(-\te)
\eeq
is satisfied and showing that the total
spin Hall conductivity is equal to zero.
 This
 is in accordance with
what has been reported in~\cite{zhang}.
Similar result has been
derived by considering a system of electrons and holes
together for a special value of
the noncommutativity parameter $\te$,
more detail can be found in~\cite{jellal2}.
Note that, thanks to (\ref{pmte}), we only use $\sigma_{xy}^s(+\te)$
in the next.

We close this  section by noting
that from the obtained results so far, it seems that
the spin down analysis is corresponding to that
for $y$-direction in spin up case and vis versa.

\section{Discussions}
Now let us turn to interpret our results. The obtained spin Hall conductivity
$\sigma_{xy}^s(+\te)$
is actually involving a free parameter $\te$. This can  be switched on to offer
different interpretations of the system under consideration. In fact,
we will show how to derive the quantum SHE
and recover three theories related to the subject.


\subsection{Quantized $\sigma_{xy}^s(+\te)$}

In the beginning,
one can notice that electrons of spin ${1\over 2}$ living on
$\mathbb{R}^2_{\te}$ behave as a SHE system
characterized by  $\sigma_{xy}^s(+\te)$.
This will be employed to establish a link with two
different theories.
The
quantum version of the intrinsic SHE can be obtained
by imposing some conditions on $\te$ and  allows us
to make contact with the Berniveg--Zhang approach~\cite{zhang}.

To talk about the quantum SHE,
we require that
$\te$ should be fixed in such way
that $\sigma_{xy}^s(+\te)$ takes quantized values in terms of
the fundamental constant ${e\over 2\pi}$.
Moreover, to get  the first quantized value of
 $\si_{xy}^s\left(+\te\right)$,
we may fix $\te$ according to
\beq
\si_{xy}^s\left(+\te\right)|_{\te=\te_{\sf bz}}=
{e\over 2\pi}
\eeq
It implies that $\te$ can be linked to the particle
density as
\beq\lb{tr}
\te={1\over \pi \rho}.
\eeq
This relation
is not surprising because it
has been derived in another formalism. Indeed, using the
noncommutative Chern-Simons theory, Susskind~\cite{susskind} showed that
to reproduce the basic features of the Laughlin theory,
for the fractional quantum Hall effect at $\nu={1\over m}$~\cite{laughlin}
resulting from charged particles, one should have $2\te ={1\over \pi \rho} $.
Moreover, if we rewrite (\ref{tr}) as
\beq\lb{surface}
\pi\te={S\over N}
\eeq
one may interpret
the quantity $\pi\te$  as an elementary surface occupied
by a quantum spin Hall droplet.
This statement is evident if we adopt
the mapping (\ref{teb}) where
the area of the quantum Hall droplet is
$2\pi l_B^2$.
Note that, the same analysis as before can be
reported for  $\sigma_{xy}^s(-\te)$.

The system under consideration is involving
 $N$-electrons of spin ${1\over 2}$. Since we have
spin up and down, one should have two sectors,
or let say two kind of particles, each one
indexed by spin up our down. If these sectors
are interacting between each other, the right wavefunctions
should be constructed in terms of the Laughlin wavefunction
analogue given before and taking into
account of the inter-correlation term
\beq
\prod_{i<j}({z}_r-\bar{w}_s)^n.
\eeq
Consequently, the required wavefunctions can be written as
\beq\lb{tot}
\Phi^m_{\sf tot}(z,\bz,\te)=
\prod_{i<j}(z_i-z_j)^m
\prod_{i<j}(\bar{w}_i-\bar{w}_j)^m\prod_{i<j}({z}_i-\bar{w}_j)^n
\exp\left[-{1\over 2\te}\left(\sum_i |z_i|^2+ |w_i|^2 \right)\right].
\eeq
They are
sharing many features with those built by Halperin~\cite{halperin}
or their equivalents in matrix model theory~\cite{jel}. The main
difference is that two different Laughlin state analogue
(\ref{mls}) and (\ref{anmls})  are resulting from
the opposite sign of the
noncommutativity parameter $\te$. 
These will be linked to those proposed for the subject,
see~\cite{zhang}.

\subsection{ Bernevig--Zhang approach}

We are wondering to prove that the present analysis is general
and can be used to reproduce other approaches,
in particular that developed by
Bernevig and Zhang~\cite{zhang}. In fact,
they quantum mechanically established a quantum theory for
the intrinsic SHE.

Let us start
by recalling that
they adopted a formalism governed
by the Hamiltonian
\beq \lb{3}
H_{\downarrow,\uparrow}=\sq {D \over 2m}
\left[p_x^2+p_y^2+x^2+y^2\pm R\left(xp_y-yp_x\right)\right]
\eeq
at a special point $R=2$ where
\beq
R={1\over 2}{C_3\over \hbar}\sqrt{2m\over D}g, \qquad
D={2mg^2C_3^2\over 16\hbar}.
\eeq
$C_3$ is a material constant,
e.g. for {\it GaAs}, ${C_3 \over\hbar}= 8\times 10^5 m/s$~\cite{dyakonov}
and g is the magnitude of the strain gradient.
This Hamiltonian is not new and was previously studied in different
contexts, one may see~\cite{howlett,pikus,bahder,khaetskii}.
It can be factorized as
\beq
 H_{\uparrow}={1\over 2\hbar}C_3g\left(2a^{\dagger}a+1\right),\qquad
 H_{\downarrow}={1\over 2\hbar}C_3g\left(2b^{\dagger}b+1\right).
\eeq
These have been used to discuss
the quantum SHE. In doing so,
Berniveg and Zhang
introduced the following configuration for the
electric field components
\beq\lb{bzef}
E_x = g x, \qquad E_y =gy
\eeq
which are analogue to what we have derived in (\ref{111}).
From their consideration, they  showed that the spin Hall
conductivity is quantized in units of $2{e\over 4\pi}$. Also
they built the corresponding wavefunctions in terms of the
Laughlin states, which sharing some common features with
Halperin ones and similar to those given by (\ref{tot}).

To reproduce the above formalism from our approach, we first
arrange our Hamiltonians in similar way as in~(\ref{3}). This can be
done by introducing the rescaling variables
\beq\lb{map}
x\lga \left(\sq{\te\over 2\hbar}\right)x,
\qquad y\lga \left(\sq{\te\over 2\hbar}\right)y
\eeq
to get a simplified Hamiltonian
\beq \lb{c}
H_{\sf hf}(\pm\te)= {m\om^2\te\over 4\hbar}  \left[
\left(p_{x}^2+p_{y}^2\right)+x^2+y^2
\pm\left(yp_{x}-xp_{y}\right)\right].
\eeq
This form is similar to that used by Berniveg and Zhang~\cite{zhang} where
the major difference is that in our approach, the term
${m\om^2\te\over 4\hbar}$ is not constant as they have.
This
suggests that our Hamiltonians are good candidates to deal with
the quantum version of the intrinsic SHE
and moreover is general in sense that one can recover other theories.

To make a link between our analysis and that reviewed above
we simply make a comparison between
our Hamiltonians~(\ref{c}) and what it is
given by~(\ref{3}). It is clear that, they have some common features.
For Berniveg and Zhang, all parameters involved in the game are constant
or depending to the material types. For us the noncommutativity parameter
$\te$ is free
and  can be fixed according
to different interpretations. Indeed, by identifying (\ref{3}) and (\ref{bzef}) to our
analogue equations (\ref{111}) and (\ref{c}),
one should choose $\te$ as $\te_{\sf bz}$ to build
a bridge between two approaches. This is
\beq
\te_{\sf bz}={ C_3g\over m\om^2},\qquad g={m\om^2\over e}.
\eeq
They lead to the constraint
\beq
\te_{\sf bz}={C_3\over e}.
\eeq
It implies that $\te$ can be interpreted as the material constant
if we set $\te =\te_{\sf bz}$
and thus can be used to characterize what kind of material is considered
to analyze the quantum SHE.
Furthermore, using
the rescaling (\ref{map}),
one can recover the wavefunctions built by Berniveg and
Zhang. The obtained derivation is proving that
our approach is relevant for the subject.

\subsection{Other theories }

The established link above is interesting in sense that
one can reproduce the Bernevig--Zhang analysis from our proposal.
This was our motivation and therefore behind the
development of our approach. But we are not going to stop at this level,
in fact
we show that how our obtained results are more generals
and one may look for other links.

To recover other theories from our proposal, we need to
introduce other mathematical tools. This can be done
by evaluating the particle number $N$ to derive another convenient
form for the spin Hall conductivity $\sigma_{xy}^s(+\te)$. Indeed, by
 definition $N$ is given by
\beq \lb{2}
N=\int_{0}^{P_F}g'd\tau(p)
\eeq
where $P_F= \hbar K_F$ is the Fermi momenta and
the quantity $g'd\tau(p)$ reads as
\beq
 g'd\tau(p)={g'S\over (2\pi\hbar)^2}d\vec{p}={g'S\over 2\pi\hbar^2}pdp.
\eeq
$g'=2s+1$ is the degeneracy,
with $s$ is the spin of particle.
Now it is easily seen that (\ref{2}) gives
\beq
N={g'S\over 2\pi\hbar^2}\int_0^{P_F}pdp={g'S\over 4\pi\hbar^2}P^2_F.
\eeq
This implies that the particle density is
\beq
\rho={g'\over 4\pi}K_F^2.
\eeq
Combining all and considering electrons of spin ${1\over 2}$
to write
the Hall conductivity for charge $\sigma_{xy}(+\te)$ along
$x$-direction as
\beq \lb{k}
\sigma_{xy}(+\te)={e^2K_F^2\over 2h}\te.
\eeq
Using the same argument as before, we obtain a spin Hall
conductivity as
\beq \lb{xx}
\sigma_{xy}^s(+\te)={eK_F^2 \over 4\pi}\te.
\eeq
This form is suggestive for our purpose and therefore
will be used to clarify our statement.
Note that
according to (\ref{pmte}),
we have an
equivalent relation to (\ref{xx})
 along $y$-direction, which is
\beq
\sigma_{yx}^s(+\te)=-{eK_F^2\over 4h}\te.
\eeq
Due to (\ref{pmte}), similar form can be found for $\sigma_{xy}^s(-\te)$
as well as that for the $y$-direction.

As we claimed before,
there are two tentatives have been theoretically elaborated to
predict the intrinsic SHE.
Among them, Sinova {\it et al.}~\cite{sinova}
who employed a mathematical formalism based on
the analysis of the  Rashba Hamiltonian
\beq\lb{rash}
H_{\sf rashba}={1\over 2m} \left(\hbar\vec K\right)^2+
\lambda(\vec \sigma \times \hbar\vec K)_z
\eeq
where $\lambda$ is the Rashba coupling constant and $\vec\si$
is the Pauli matrix.
 They showed that the spin Hall conductivity
has an universal value, such as
\beq\lb{sinova}
\sigma_{\sf rashba}^s={e\over 8\pi}.
\eeq
Subsequently, by considering the Rashba--Dresselhaus spin--orbit
coupling, it shown that (\ref{sinova}) can be generalized
to~\cite{sinova2,shen}
\beq
\sigma_{\sf rashba-dress}^s=\pm{e\over 8\pi}.
\eeq
We mention that a related work has been reported by Hu~\cite{hu}
in analyzing the topological orbital angular momentum Hall
current.
This issue will be considered separately in a forthcoming paper.

To reproduce the Sinova {\it et al.} analysis, we solve the
equation
\beq
\sigma_{xy}^s\left(+\te\right)|_{\te=\te_{\sf rashba}} = \sigma_{\sf rashba}^s
\eeq
to get a fixed noncommutativity parameter
\beq
\te_{\sf rashba}={1\over 2K_F^2}.
\eeq
Therefore, one can  envisage electrons of spin ${1\over 2}$
living on ${\mathbb R}^2_{\te}$
as the Rashba system described by (\ref{rash}) if we
set $\te=\te_{\sf rashba}$.

On the other hand, Murakami {\it et al.} proposed an interesting approach
to predict the intrinsic SHE~\cite{murakami1},
see also~\cite{murakami2}. Their analysis was
based on the investigation of the basic features of the
 Luttinger Hamiltonian given by
\beq
H_{\sf luttinger}={\hbar^2 \over 2m}\left[\left(\gamma_1+{5\over2}
\gamma_2\right)K^2-2\gamma_2\left(\vec{K}.\vec{S}\right)^2\right]
\eeq
where $\gamma_1$ and $\gamma_2$ are the valence-band parameters for
semiconductor materials.
This form of $H_{\sf luttinger}$ allowed them to describe
the phenomena by showing that
the spin Hall conductivity for heavy and light holes can be written as
\beq
\si_{\sf luttinger}^s={e\over6\pi^2}\left(3K_F^H-K_F^L\right).
\eeq

Clearly, to reproduce the Murakami {\it et al.} results, we first swish our system to
that of holes. Simply this can be done by changing in our analysis $e$
by $-e$ to get $-\sigma_{xy}^s(+\te)$. Therefore, requiring that
the identification is satisfied
  \beq
-\sigma_{xy}^s(+\te)|_{\te=\te_{\sf luttinger}} = \sigma_{\sf luttinger}^s
\eeq
we end up with the condition on $\te$
\beq
\te_{\sf luttinger}={2\over 3\pi K_F^2}\left(K_F^L-3K_F^H\right).
\eeq
Our analysis offered for us two possibilities to talk about
the intrinsic SHE. Semi-classically, fixing $\te$ we have made a
connection to what
have been reported by  Sinova {\it et al.} and Murakami {\it et al.}
on the subject.
Quantum mechanically, we have reproduced the
Berniveg--Zhang analysis where the quantized spin
Hall conductivity and the corresponding wavefunctions have been
identified.

\section {Conclusion}

To discuss the intrinsic spin Hall effect, we have employed two noncommutative
harmonic oscillators
and investigated their basic features.
Indeed, restricting
to the high frequency regime, we have derived a factorized Hamiltonians
analogue to those have been used by  Bernevig and Zhang~\cite{zhang} in dealing with
the quantum spin Hall effect. Moreover, we have shown its
 common features with the Landau
problem on the ordinary plane. Getting the spectrum for
spin up and down cases, we have determined the
spin Hall conductivities $\sigma_{xy}^s(\pm\te)$
in a general form due to the
 noncommutativity
parameter $\te$-dependency. Moreover, they have been obtained without
need of an external magnetic field and showing that
the total spin Hall conductivity is null.

Subsequently, by fixing $\te$ differently, we
have given some discussions. Indeed,
to get a quantum version of the intrinsic
 spin Hall effect, we have required that
the obtained $\sigma_{xy}^s(\pm\te)$
should be quantized in terms of the fundamental constant ${e\over 2\pi}$.
The corresponding wavefunctions have been constructed in similar way
as those built by Halperin. This interpretation offered for us a possibility
to make contact with the Berniving--Zhang approach~\cite{zhang}. In fact
by choosing a particular value of $\te$, we have noticed that $\te$
can be used to determine what kind of material is
considered to analyze the subject
and therefore reproduced the Berniving--Zhang analysis.

Evaluating the particle number in terms of
Fermi momenta, we have derived another form of
the conductivities $\sigma_{xy}^s(+\te)$
in terms of the Fermi wave vector. This allowed us to establish other links
with different approaches. Indeed, giving to $\te$ two different
values, we have shown that the Sinova {\it et al.}  and Murakami {\it et al.}
analysis can be recovered from our proposal.

Still some interesting questions to be addressed. Can we use the noncommutative
Chern-Simons theory~\cite{susskind,poly} to describe the basic features of
the quantum spin Hall effect? A related question arose, in fact
what about a matrix model description of the phenomena? These issues
and related matters are under consideration.

\section*{Acknowledgments}

The authors are thankful to R. El Moznine for
fruitful discussions on the high frequency regime.
AJ~work's was partially
supported by Arab Regional Fellows Program (ARFP) $2006/2007$.

\end{document}